# In-The-Wild Interference Characterization and Modelling for Electro-Quasistatic-HBC with Miniaturized Wearables

Parikha Mehrotra\*, David Yang\*, Student IEEE member, Scott Weigand† and Shreyas Sen\*, Senior *Member, IEEE*

\**School of Electrical and Computer Engineering, Purdue University* †Eli Lilly and Company

*Abstract—* The emergence of Human Body Communication (HBC) as an alternative to wireless body area networks (WBAN) has led to the development of small sized, energy efficient and more secure wearable and implantable devices forming a network in and around the body. Previous studies claim that though HBC is comparatively more secure than WBAN, nevertheless, the electromagnetic (EM) radiative nature of HBC in >10MHz region makes the information susceptible to eavesdropping. Furthermore, interferences may be picked up by the body due to the human body antenna effect in the 40-400MHz range. Alternatively, electro-quasistatic (EQS) mode of HBC forms an attractive way for covert data transmission in the sub 10MHz region by allowing the signal to be contained within the body. However, there is a gap in the knowledge about the mechanism and sources of interference in this region (crucial in allowing for proper choice of data transmission band). In this paper, the interference coupling modality in the EQS region is explained along with its possible sources. Interferences seen by the wearable in the actual scenario is a non-trivial problem and a suitable measurement EQS HBC setup is designed to recreate it by employing a wearable sized measurement setup having a small ground plane. For the first time, a human biophysical interference pickup model is proposed and interference measurement results using a wearable device are presented up to 250kHz in different environmental settings.

## I. INTRODUCTION

The advent of technological evolution has led to the development of miniaturized and small form-factor on-body internet connected devices [1]. This introduction of small-sized wireless body area network (WBAN) wearables, injectables, ingestibles and implantables [2] has transformed the fields of healthcare monitoring and secure data transmission, by enabling information exchange to biomedical sensors, smart watches, mobile phones and allowed remote and continuous patient monitoring and feedback. Typically, WBAN devices employ radio waves to wirelessly communicate information. However, being radiative in nature they not only require a high power for transmission as they attenuate in power density while propagating through space but also can be easily intercepted by malicious eavesdroppers and get access to private user information [3]. Thus, there is need for more energy efficient and secure data exchange techniques. Alternatively, human body having a high-water content, and acting as a lossy wire-like conductive channel for signal propagation can be utilized [4], [5], [3], [6], [7]. EM mode of HBC (>10MHz) has predominantly been used for the purpose of communication with the primary bottleneck being the interference picked up by the human body antenna effect in the 40M-400MHz range [8]–[10]. To avert this, narrowband (NB) implementations (20-80MHz carrier frequency) utilizing static frequency bands with tolerable interference or adaptively hopping between bands based on the measured channel quality have become popularized [11], [12]. NB HBC, couples the modulated narrowband EM signals to the human body using a coupler instead of radiating it with an antenna. This reduces the energy consumption of 10nJ/bit in conventional WBAN systems to 110pJ/bit in NB HBC [3]. Although, NB HBC is more energy efficient than WBAN, the EM nature of communication nevertheless radiates signal outside the body rendering it as not secure for data transmission. Electro-quasistatic (EQS) mode of signal transmission using HBC forms an attractive means for energy efficient and covert data exchange for sub 10MHz frequencies [3] as it couples the signal through the conductive layers below the skin enabling containment of the signal and has been shown to exhibit energy efficiency up to 10pJ/bit [5]. However, to ensure efficient communication via EQS HBC, it is essential to characterize and model the interferences present. This paper explores the mechanism and sources of interferences by performing in-the-wild measurements using a wearable sized measurement device to emulate the interference picked up by a real scenario EQS wearable receiver.

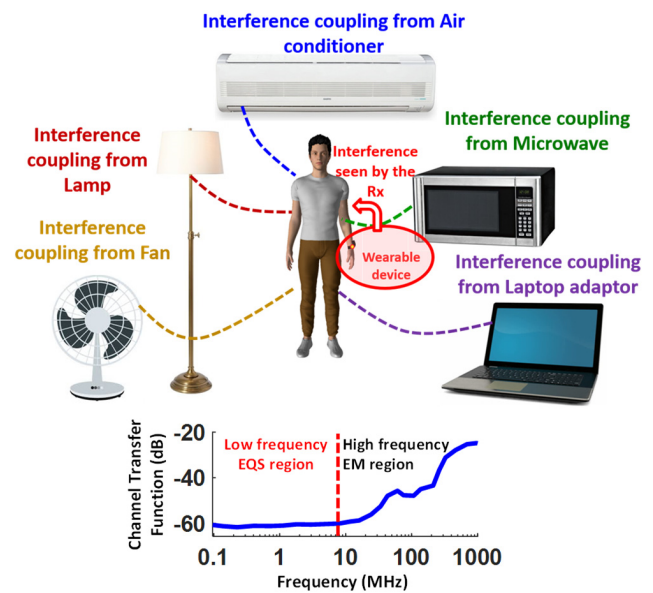

Figure 1. Environmental interferences from electrical devices such as lamps, fans, and laptops couple to the human body posing a limitation on its effective usage as a communication channel in the EQS region. The plot shows the human body as a low loss broadband channel in the EQS region (sub 10MHz).

This paper is divided into the following sections: section I builds insight into using EQS HBC as an alternative to WBAN to achieve better energy efficiency and security, in section II we discuss the motivation for analyzing the environmental interferences which may hamper EQS HBC as well as discuss the need for a wearable measurement device, section III gives a brief inkling of the previous HBC works and how we circumvent their shortcomings. A novel biophysical model for the human-interference pickup is elucidated in section IV followed by the design of a wearable sized interference measurement device using COTS components in section V. Section VI highlights the important considerations while analyzing the interference measurement results to distinguish between actual environmental interferences and peaks from non-idealities, followed by wearable device's in-the-wild measurement results in section VII and finally, the paper is concluded in Section VIII.

## II. Motivation

To effectively utilize the human body as a communication channel in the EQS region, it is desirable to understand the existing interferences and its sources in this frequency band (<10MHz). However, it is a non-trivial problem as there is gap in the knowledge about the mechanism of interference coupling and its sources in the EQS HBC region. Additionally, the optimal EQS HBC scenario entails wearable devices having high capacitive input impedance and floating ground planes, hence, to correctly determine the interference magnitude picked up by these devices, grounded equipment previously used in HBC studies such as oscilloscopes (OSC) and spectrum analyzers (SA) become ineffective.

### A. *Need for understanding interferences in the EQS region*

In previous studies, human body was characterized as a high pass channel having high loss at low frequencies making these frequencies unusable [4] and mostly focused on the EM mode for HBC (>10MHz). However recently, using voltage mode signaling and high impedance capacitive termination at the receiver end, the human body channel has shown to provide flat-band loss at low frequencies (Figure 1) [7], [13] with the primary bottleneck being the effect of interference picked up by the body [10]. Thus, knowing the mechanism and effects of interference in the sub 10MHz frequencies is of great importance for effectively using EQS HBC for secure and energy efficient data exchange.

There have been studies conducted on the effects and mechanisms of interference picked up by the human body only in the EM HBC region [14], where the interference coupling mechanism is through the human body antenna effect. However, in the EQS region (at low frequencies ~100kHz), the antenna size (in this case the human body) needs to be around 750m to act as an efficient antenna. Alternatively, in EQS HBC mode, the interference picked up by the human body can be explained through capacitive coupling [15]. In this mode, the signal from a transmitter gets capacitively coupled to the surface of human body creating electric fields and the receiver picks up the displacement current. The return path can be modelled as a capacitance between the earth's and the wearable device's ground [16]. Large surface area radiating devices act as interference sources and get capacitively coupled to the skin with nearby radiating sources having more impact on the received interference signal as compared to those at a greater distance.

### B. *Need for a miniaturized wearable device for EQS HBC interference measurement*

In EQS HBC, the device's transmitting/receiving electrodes are connected to the human body while the ground electrodes are left floating such that the human body forms the forward path of communication and the return path is formed by parasitic capacitances between the environment (earth ground) and the transmitting/receiving ground planes. The EQS HBC wearable devices are typically made with CMOS technology and have high capacitive input impedance with their ground planes floating. To effectively design EQS HBC transceivers, one must be mindful of the existing environmental interferences' magnitude and frequency. This received interference magnitude depends on the receiver's input impedance; hence it is important that the measuring equipment used must replicate the EQS receiver's input impedance. Furthermore, the return path capacitance for a floating ground wearable receiver is quite different from that of an earth grounded device, impacting the magnitude of interference observed. Both SA as well as OSC fail to satisfy these requirements. Figure 2 intuitively compares the three scenarios for measuring the interference coupled to the body using a SA, OSC and a wearable (WR) measurement device. The SA is a small, resistive, 50Ω terminated grounded device (having a large ground plane), while the OSC is a high input capacitive impedance terminated grounded device (also having a large ground plane). A small resistive termination implies a lower received signal than expected in an EQS HBC scenario, while a large ground plane implies receiving a higher magnitude of interference voltage than expected, leading to pessimistic estimations of received interference. Hence, to closely match the magnitude of the interferences picked up by a wearable receiver, a miniaturized wearable interference measurement device is needed, both to match the input impedance of the actual wearable receiver as well as a floating ground condition (a small ground plane).

## III. Literature survey

In previous studies [8], with the goal of determining the frequency bands with maximum HBC gain, the human subject was exposed to EM fields by intentionally applying a signal from a battery powered RF source generating 1-200MHz (i.e. EM HBC region) signal through a 60cm monopole antenna located 3m away as summarized in Figure 3. Human body and human arm resonances, i.e. maximum HBC gain frequency bands were estimated using an AC-powered SA (low input impedance grounded device) connected to an electrode touched by the human subject. In another study, human body antenna effect in the EM region of 30-120MHz was observed using a 4-channel adaptive frequency hopping scheme [11]. The in-band interferences coupled to the body were measured on a SA (as seen earlier gives pessimistic interference measurement results) through a coaxial cable and a metal electrode attached to the left hand. The interferences through cordless phones and walkie-talkies were intentionally generated and coupled to the body and visualized. A 1.5Vpp signal at the transmitter side got attenuated to -50dBm by the 1.8m body channel. The reduction in signal to interference ratio (SIR) in the FM band (80-110MHz) was up to -20dB, while the walkie talkie degraded the SIR to -22dB. Both mentioned works studied the EM HBC behavior using intentional signals measured using earth grounded devices. Whereas, our focus in this paper is to examine the not so well known existing environmental interferences that could interfere with EQS HBC (sub 10MHz) signals on getting capacitively coupled to the body. This is performed by using a miniaturized wearable measurement device as opposed to grounded devices used in previous studies that do not replicate actual wearable EQS HBC scenario.

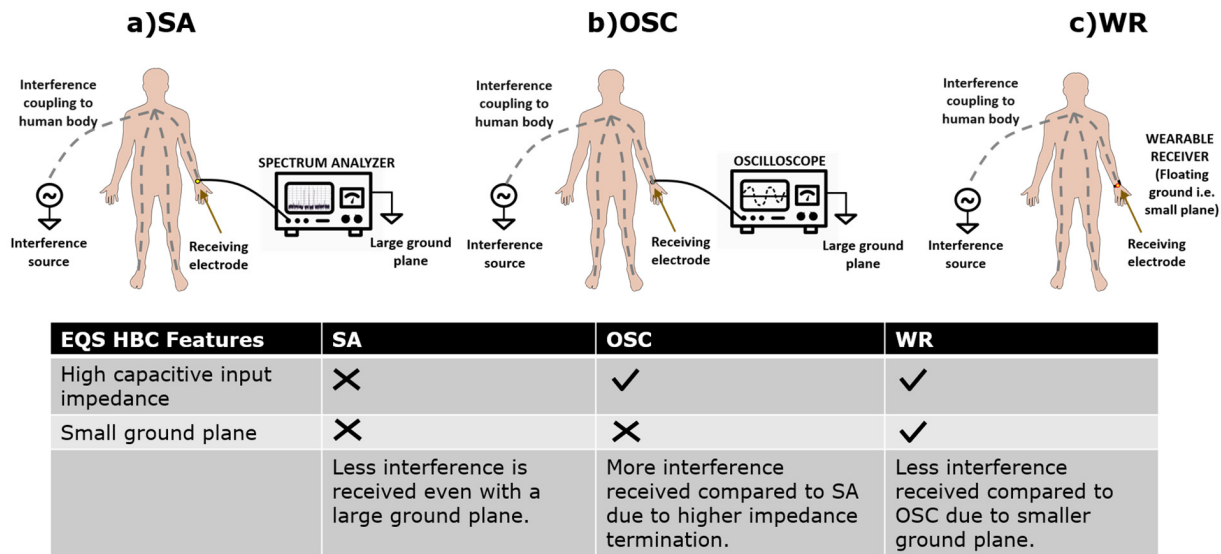

Figure 2. Choice of proper EQS HBC interference measurement setup which most closely replicates the interference picked up in the actual wearable scenario.

IV. HUMAN BIOPHYSICAL MODEL FOR INTERFERENCE PICKUP

As mentioned earlier, the goal of this paper is to determine the existing environmental interferences that may couple to the body and degrade the performance of the EQS HBC wearable receiver. In this section, we will quantitatively distinguish between the three devices using circuit models; namely the SA, OSC and WR and determine how well they replicate the EQS HBC scenario. However, before we delve into the choice of equipment to be used for interference measurement, it is crucial to understand the passive models of the human body and of interference coupling to the human body as shown in Figure 4. An interference source is essentially an EM radiation source, whose electric component couples to the body capacitively through $C_{INTF}$, assuming the body to be an equipotential surface. The interference source can have a floating ground and, in that case, introduces an interference source return path capacitance, $C_{INTF-GND}$. The coupled signal suffers impedance by the body tissue and skin as it flows through the body, denoted as $Z_{BODY}$ (~1-10kΩ consisting of tissue resistance in series with parallel combination of skin resistance and capacitance) [7], [17], [18]. This is received by the body worn wearable receiver through a tight skin contact with the receiver electrode ($C_{BAND}$ (~200pF), $R_{BAND}$ (~100Ω)) followed by the receiver's input termination (i.e. SA/OSC/WR) denoted by $C_{RX}$ and $R_{RX}$. Return path capacitances exist between the body and earth ground at the receiving end ($C_{BODY}$ of ~150pF) and between the receiver ground and earth ground (when the receiver has a floating ground, $C_G$ is ~1.2pF, else can be considered a short circuit).

A simplified circuit diagram of human body interference pickup is shown in Figure 5. Two possible cases are considered, namely, when the interference source is grounded and when the interference source is floating. In the latter, an extra $C_{INTF-GND}$ return path capacitance gets added in series with the $C_{INTF}$. $V_{INTF}$ is the input interference voltage that gets coupled to the body and $V_{WR}$ represents the voltage received by the wearable. The forward path impedance ($Z_{BODY}$) is neglected in this calculation, as $Z_{BODY}$ is of the order of 1-10k$\Omega$, whereas the impedance provided by $C_{INTF}$ is assumed greater than 100k$\Omega$, for sub-MHz frequencies. Furthermore, the return path capacitance ($C_G$) is an order of magnitude smaller than the load capacitance ($C_{RX}$). Using these approximations, we obtain the ratio of $V_{WR}$ and $V_{INTF}$ as a product of two terms as shown in Figure 5. The first term shows how much interference gets coupled to the body and is a function of $C_{INTF}$ and $C_{INTF-GND}$. Thus, the first term reduces with increase in distance between the body and the interference source as it reduces the capacitive coupling ($C_{INTF}$) between them. It is further reduced, when the interference sources are floating due to the incorporation of $C_{INTF-GND}$, reducing the numerator of the first term. The second term depicts the percentage of the coupled voltage picked up by the wearable and is a function of the wearable geometry due to its dependence on $C_G$ [18].

| Parameter | [8] | [11] | Our work |
|---|---|---|---|
| Frequency Range | 1-200MHz | 30-120MHz | 1-250kHz |
| Source of signal | Monopole antenna using battery powered RF source | Antenna using battery-powered signal generator | 'In-The-Wild' Environmental interference |
| 'In-The-Wild' | No | No | Yes |
| Antenna / Capacitor coupling | Antenna | Antenna | Capacitor |
| Measurement device | Spectrum analyzer | Spectrum analyzer | Miniaturized wearable |
| Input impedance of device | 50Ω resistive impedance | 50Ω resistive impedance | KΩ capacitive impedance |
| Grounded/ non-grounded | Grounded | Grounded | Non-grounded |
| Goal | Maximize HBC gain in EM HBC region using intentional signals | Determining SIR in EM HBC region using intentional signals | Determining unknown environmental interference in EQS HBC region |

Figure 3. Comparison summary between previous works and this paper.

Having developed an understanding of the biophysical model for interference pickup, we can modify it for different receiving equipment as shown in Figure 6. With Spectrum Analyzer as the measurement device for picking up interference coupled to the body, the load offered by it is resistive and small in impedance (i.e. $R_{SA}$ is 50$\Omega$). Also, since it's plugged to the power supply ground, $C_G$ is short circuited. This implies that most of the signal is lost due to the high impedance of the body and the received interference signal is lower than what is expected in a wearable scenario. Assuming $C_{INTF}$ to be 30pF, the normalized received voltage at the SA ($V_{SA}/V_{INTF}$) has a frequency dependence and ranges from -53 to -100dB in the 1-250kHz EQS region (Figure 6a). In case of OSC, the termination impedance is a parallel combination of $C_O$ (80pF) and $R_O$ (1M$\Omega$) [7]. Due to both its input impedance being high and being directly connected to the power ground, a higher interference voltage is received than would be expected in a wearable EQS scenario. Figure 6b shows the normalized received voltage at the OSC to be ($V_{OSC}/V_{INTF}$) -18dB. The input impedance of the wearable device measured using an LCR meter is ($C_L$) 20pF. Connecting the wearable receiver to ground (referred as WRG in Figure 6c, where $C_G$ gets short circuited), it behaves like the OSC, both having high capacitive

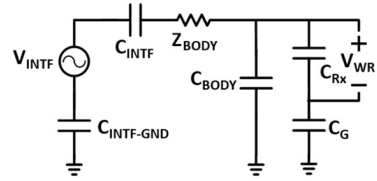

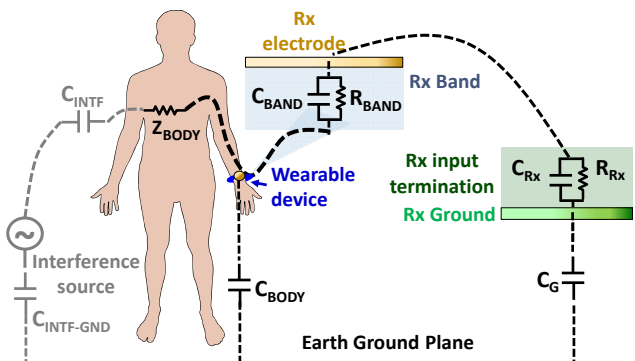

Figure 4. Biophysical model of the human body capacitively coupled to a floating interference source.

Figure 5. Circuit diagram of the human body capacitively coupled to an interference source.

input impedance and earthed ground. The normalized received voltage ($V_{WRG}/V_{INTF}$ = -16dB) is therefore very close to that of OSC. Finally, in Figure 6d, the wearable ground is left floating, providing a $C_G$ comparable to that expected in an EQS HBC receiver scenario and receives a normalized voltage of ($V_{WR}/V_{INTF}$) -38dB. From this analysis, we can approximate the variation of received interference using different measuring instruments: SA receives ~15 to 62dB less voltage than expected in EQS scenario, while an OSC receives ~22dB more. This motivates the requirement for a miniaturized battery powered measurement device for interference measurement purposes as both the previously used devices, SA and OSC

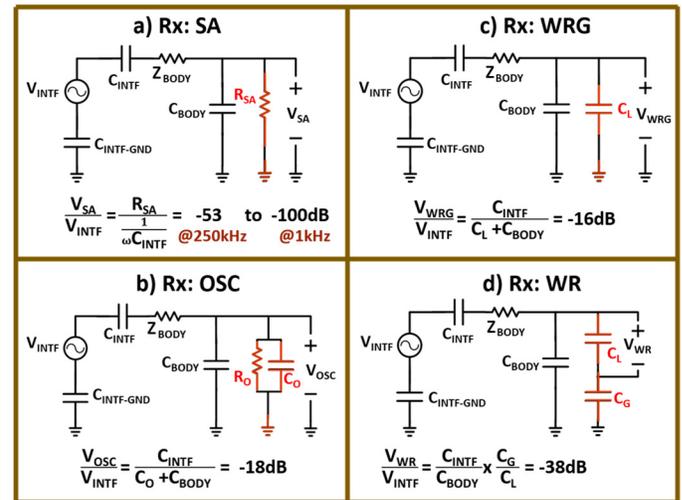

Figure 6. Circuit diagram of the human body capacitively coupled to an interference source and measured using different devices (SA, OSC, WRG, WR).

cannot be used to accurately determine the real scenario EQS HBC interference magnitude.

V. DESIGN OF MINIATURIZED WEARABLE MEASUREMENT DEVICE

The interference signals get coupled to the body and are received by the wearable electrode. The conceptual block diagram of this interference measurement setup along with the actual assembly using commercial off the shelf components (COTS) is shown in Figure 7. These signals go through a buffer, placed at the input to provide a high impedance termination. They are then amplified for better detectability by the ADC followed by a low pass anti-aliasing filter with a cutoff at 250kHz. The received signal being a voltage signal is sampled in time domain for measurement. For this purpose, an ADC is employed to sample and digitize the interference affected received waveform. TM4C123GH6PM micro-controller is used having a 12-bit ADC resolution and a 500kHz sampling rate. The sampled data is stored in the micro-controller which forms a part of the receiving wearable device and is transmitted through the Bluetooth module HC05 to another HC05 module connected to a PC for post processing. Through this we obtain the interference FFT spectrum in the EQS region up to 250kHz.

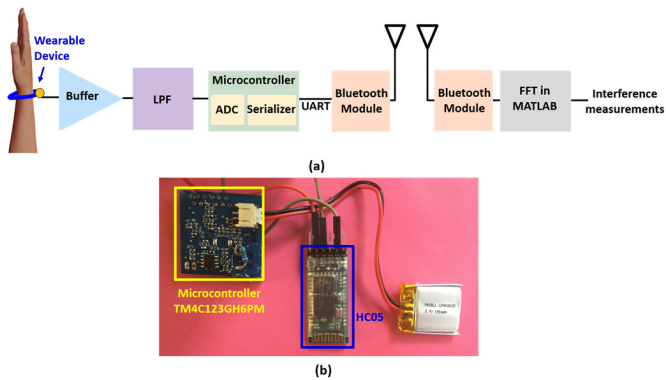

Figure 7. (a) Block diagram of the wearable interference measurement setup, (b) Wearable setup using COTS components to acquire the interference data.

The need for high impedance termination in the EQS receiver is to achieve low channel loss through the body. Figure 8a shows the EQS HBC scenario where data is sent through a wearable transmitter, suffers impedance by the body and is received by the body worn wearable receiver. In this case, return path capacitances exist between the body and earth ground ($C_{BODY}$ of ~150pF) and between the wearable grounds and earth ground ($C_{GTX}$, $C_{GRX}$ ~1.2pF). Figure 8b shows the simplified circuit model of EQS HBC and it can be noted that to minimize channel loss through the body (or in other words, to increase the received voltage $V_{WR}$), the design choice is to minimize $C_{RX}$ (as increasing $C_{GRX}$ would imply increasing the wearable device's ground plane size [18]). This is achieved by placing a buffer at the wearable receiver's input which provides a $C_{RX}$ ~20pF.

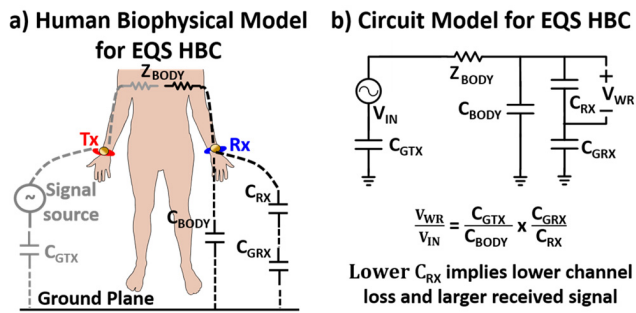

Figure 8. a) Human biophysical model for EQS HBC scenario, b) A simplified circuit model explaining the choice for a high capacitive input impedance EQS HBC wearable receiver for minimizing channel loss.

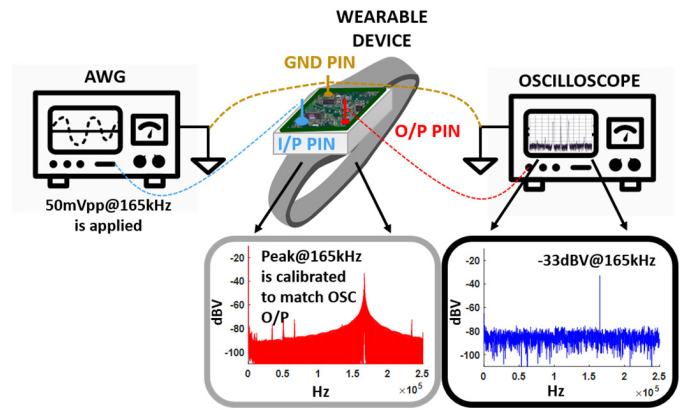

Figure 9. Calibration setup: A signal is applied using an AWG and the received voltage at the output pin in the wearable is probed by an oscilloscope. This probed voltage value is matched to the wearable device's output value during post processing.

**Calibration:** The wearable measurement device is calibrated using the setup as shown in Figure 9. A known signal is applied using an arbitrary waveform generator (AWG) and connected to the miniaturized wearable receiver's input electrode. The amplified signal at the ADC input is probed using an OSC and matched to the AWG applied signal's amplitude as seen in the post processed FFT spectrum. It is to be noted, that calibration is carried out with the wearable device's ground earthed (i.e. on WRG instead of WR). This is advisable when using strong grounded AWG and OSC, as keeping the wearable ground pin floating may create ambiguities coming from the return path ($C_G$) and affect the accuracy of calibration.

## VI. IMPORTANT MEASUREMENT CONSIDERATIONS

The goal of this paper is to identify the existing environmental interferences in the EQS region (<250kHz), which may come from devices encountered in our everyday lives such as lights, fans, air conditioners etc. and may interfere with EQS HBC. However, in order to correctly interpret the information obtained from the wearable measurement device, it is essential to understand the inherent nature of the wearable device first (i.e. to figure out if the device itself can add peaks to the spectrum) as well as the different phenomena and second order effects such as mixing and buffer nonlinearities which may create peaks in the frequency spectrum that may not be from actual environmental interferences. Also, developing an understanding of the noise floor (NF) and its dependencies is essential in characterizing the signal to noise ratio (SNR) and signal to interference ratio (SIR) for EQS HBC. Many of these challenges are unique to high impedance microcontroller-based measurement systems which have not been addressed in literature before but is the only correct way and needs a detailed discussion.

### A. Inherent Device peaks

To determine if the wearable measurement device itself caused any interference peaks, the EQS spectrum was observed in an anechoic chamber. On plotting the spectrum up to 250kHz in Figure 10 using grounded wearable device (WRG) with a sampling rate of 500kHz (controlled by the ADC in the microcontroller using reference clock of 80MHz), peaks @50kHz, 100kHz, 150kHz and 200kHz were observed. The only possible source that could create those peaks being in an anechoic chamber was the device itself. A second confirmation that the device was generating these peaks was achieved by changing the sampling rate to 450kHz which shifted the peaks to 45kHz, 90kHz, 135kHz and 180kHz. If the peaks were from any other interference source, they would have retained their positions (i.e. we would have still seen peaks @50k, 100k, 150k and 200kHz) even after changing the sampling rate. Another, indicator that these peaks come from the wearable device is that the peak magnitudes remain unchanged in the WRG and WR measurements, despite the additional factor from capacitive division due to the return path capacitance. Thus, it can be concluded that the wearable itself creates peaks in the spectrum (due to the microcontroller generated clock), depending on the sampling rate and are not to be confused with any interference source in further analysis.

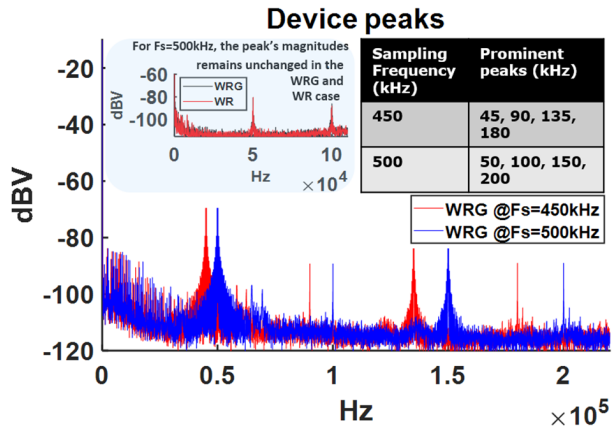

Figure 10. Device peaks observed when sources of interference are turned off and sampling rates are varied (450k, 500kHz). The subplot shows that the peak magnitude of the interferences (for Fs = 500kHz) remains unchanged in the WRG and WR case.

B. *Effect of non-ideal anti-aliasing filter*

An anti-aliasing low pass filter is added in the signal path of the wearable device such that any interference signal greater than the maximum frequency (= sampling rate/2 ~250kHz) gets filtered out. However, due to the non-ideality of the second order LC filter employed due to size and filer order limitation, the rejection obtained beyond the cutoff frequency (>250kHz) is moderate. Due to this, interference signals >250kHz fold back and appear as peaks in the spectrum (following the equation; $f_a = f \pm kF_s$ where $f_a$ is the aliased frequency, $F_s$ is the sampling rate and k is an integer). This effect was verified by applying an intentional signal of 740kHz from an AWG on WRG and was seen as an alias @240kHz in Figure 11.

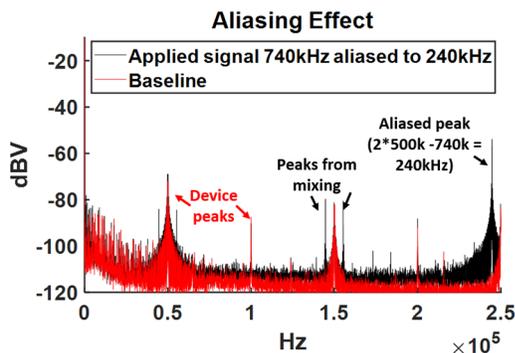

Figure 11. On applying a signal (740kHz) greater than the sampling rate (500kHz), an alias is observed (@240kHz) due to the non-ideality of the antialiasing filter (with cutoff @250kHz).

C. *Effect of mixing of interferences*

Second order effects of mixing were observed while using the wearable device, where the interference signals got mixed with the device generated peaks (50kHz, 100kHz, 150kHz, 200kHz) to create new peaks. This was verified by applying a 240kHz signal from an AWG on WRG as shown in Figure 12. On mixing with the device's peaks- 100kHz, and 200kHz (with $F_s$=500kHz), generated new peaks @ 40kHz, 60kHz, 140kHz and 160kHz. The peaks at 60kHz and 160kHz were aliased versions of 340kHz and 440kHz as shown in Figure 12's table, while the peaks generated from mixing between 240kHz applied signal and the device peaks- 50kHz and 150kHz were smaller in magnitude and remained below the noise floor.

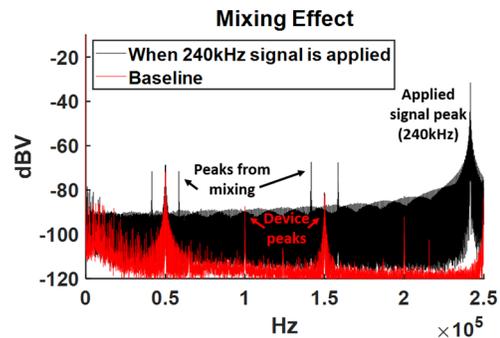

| Peak 1 from applied signal (kHz) | Peak 2 from device when Fs=500ksps (kHz) | Mixed Peaks (Peak1±Peak2) | Comments |
|---|---|---|---|
| 240 | 50 | 290 $\xrightarrow{aliased\ to}$ 210, 190 | Under NF |
| 240 | 100 | 340 $\xrightarrow{aliased\ to}$ 160, 140 | Visible in Fig. |
| 240 | 150 | 390 $\xrightarrow{aliased\ to}$ 110, 90 | Under NF |
| 240 | 200 | 440 $\xrightarrow{aliased\ to}$ 60, 40 | Visible in Fig. |

Figure 12. The applied 240kHz signal gets mixed with the device peaks (50k, 100k, 150k and 200kHz) to generate new peaks @ 40k, 60k,140k and 160kHz.

D. *Effect of buffer Non-linearities*

Another second order effect observed in the wearable measurement device's spectrum was from buffer nonlinearity creating new peaks at integer multiples of the interference signal. To illustrate this effect, a 44kHz signal was applied from an AWG to WRG as shown in Figure 13, and multiples of this signal were seen in the spectrum, i.e. @ 88kHz (2[nd] order) and 132kHz (3[rd] order). The higher order multiples reduced in amplitude and got buried under the noise floor. Thus, with respect to the baseline (where only device generated peaks are observed), on applying an intentional signal, all the additional peaks created can be explained through the combined effects of mixing and buffer nonidealities.

E. *Effects of sample size, FFT size, signal strength and windowing on the Noise floor*

The sensitivity of the wearable depends on the value of noise floor (NF), as any signal with a magnitude below it will not get detected, thus it is important to keep it to a minimum. It is possible to control the NF by choosing a proper duration (L) for which the ADC samples are recorded, and by choosing proper number of frequency samples (i.e. FFT points 'N') during FFT as shown in Figure 14a & b. From Figure 14a it can be observed that on increasing the number of ADC digitized samples (i.e. feeding a longer length of samples to the post processing FFT algorithm), from L = 0.1M to 1M the NF reduces from -112dBV to -120dBV. Similarly, on increasing the number of frequency samples during FFT from N= $2^8$ to $2^{16}$, the NF reduces from -90dBV to -110dBV (Figure 14b). Another factor which influences NF is spectral leakage, which occurs when the sample frequency is not an integral multiple of the FFT resolution, i.e. (Fs/N) and causes leakage of energy to other frequency bins as well instead of just the nearest frequency bin value [19]. Due to the unknown nature of the interferences being measured, it is not feasible to avert this effect. As a result, we encounter signal amplitude dependence of NF, as leakage of the signal to surrounding bins also increases with the increase in signal amplitude. Figure 14c illustrates this by applying a known amplitude signal from the AWG (12.5mVpp, 25mVpp, 50mVpp and 100mVpp) and the NF is seen to increase by 20dBV with a 4X increase in signal amplitude. Spectral leakage can be reduced by choosing an appropriate window function as shown in Figure 14d which illustrates the effect of different windowing techniques (Hann, Blackman and Chebwin) on NF. Taking into consideration all the above factors, L of $10^5$ samples and N of $2^{16}$ is chosen in our FFT algorithm for taking interference measurements with a suitable NF.

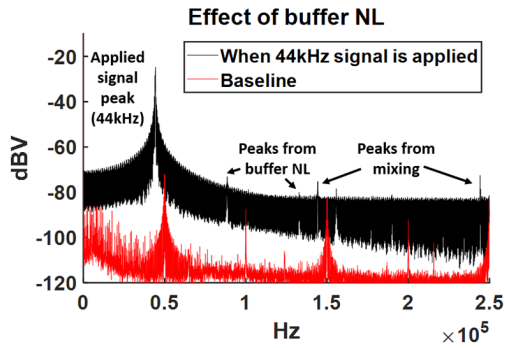

| Peaks from Buffer NL (kHz) | Comments | Mixed Peaks (kHz) | Comments |
|---|---|---|---|
| 44x2=88 | Visible in Fig. | 44±50=94, 6 | Under NF |
| 44x3=132 | Visible in Fig. | 44±100=144, 56 | Visible in Fig. |
| 44x4=176 | Under NF | 44±150=194, 106 | Under NF |
| 44x5=220 | Under NF | 44±200=244, 156 | Visible in Fig. |

Figure 13. The applied 44kHz signal creates new peaks due to buffer nonlinearities which are integer multiples of the applied signal.

F. *Received signal magnitude in an oscilloscope, grounded wearable device and non-grounded wearable device*

So far, we have applied AWG signals directly to the wearable receiver for verifying second order effects and to develop an understanding of the noise floor dependencies. To validate the theoretical models of Figure 6 with different measuring devices, an AWG signal is applied on the body using a low impedance probe and is measured using an OSC, WRG and WR (Figure 15). As predicted by the models in Figure 6, the OSC and WRG received values are similar (-43dBV @170kHz), while the peak from WR is -13dBV lower (-56dBV @170kHz). Apart from the intentional applied signal, we also observe an interference signal at 215kHz in Figure 15. The received signal amplitude in WRG is -50dBV, while it is -60dBV in WR, which is again a -10dBV drop as expected.

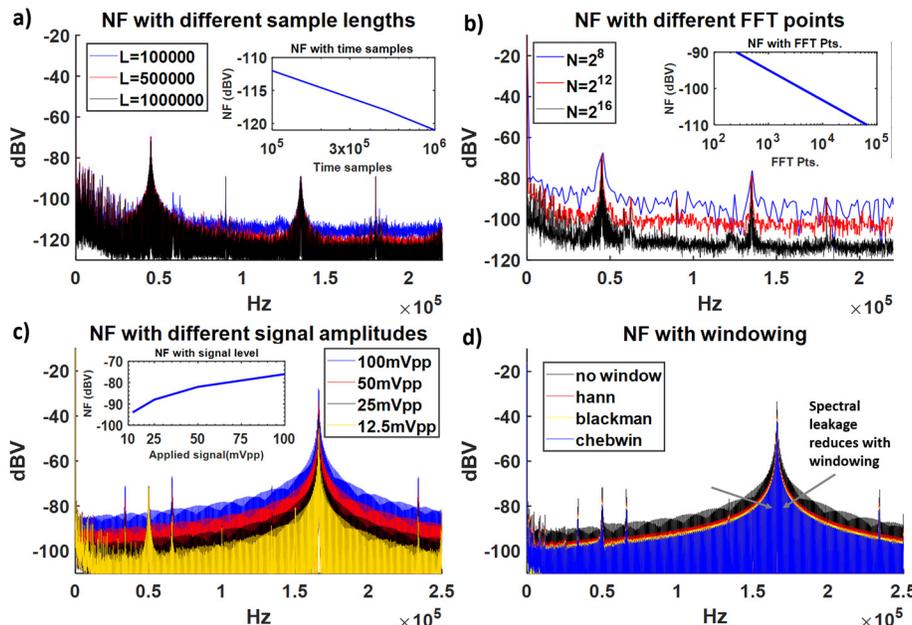

Figure 14. Variation in value of noise floor (NF) with a) different lengths of digitized ADC samples, b) different number of FFT points/ bin size, c) different applied signal amplitudes, d) different windows to reduce spectral leakage during FFT.

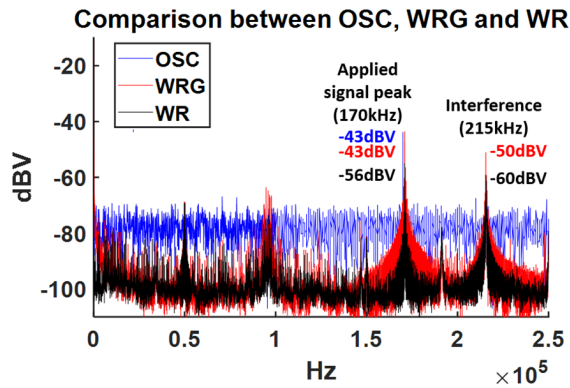

Figure 15. A 170kHz signal is applied from AWG and measured using OSC, WRG and WR.

## VII. IN-THE-WILD MEASUREMENT RESULTS

With this understanding about the different measurement devices available for interference measurement, we can now figure out the sources, the correct signal magnitudes and the frequencies of possible interferences which may exist in the EQS region (<250kHz). For this purpose, we first collect baseline data at different locations; namely, a house bedroom, living room, an equipped electronic lab and a commercial building hall (Figure 16a). The house setting being a more flexible environment, the baseline is taken with all electronic appliances (such as lights, fans, refrigerators, stoves) plugged off, while the baseline in the lab is only with the lights turned off and all other appliances on (including computers, scopes, and signal generators). The baseline in the commercial building hall is without plugging out any appliance (and includes ceiling lights and a television). Measurements are carried out using an OSC, WRG and WR (where only the WR represents the actual wearable EQS HBC scenario). Once the baselines are acquired, the individual appliances can be turned on one at a time to observe their individual effect (i.e. whether they are responsible for any interference peaks in the EQS region) on the three measurement devices. The appliances tested for interference include – an air conditioner, a fan, CFL light lamps, LED light lamps, fluorescent tubes, flush mounted CFL lights, a microwave, dimmable fluorescent tube lights, a stove, a refrigerator, a washing machine and a laptop adaptor. On taking measurements after turning the bedroom air conditioner on (Figure 17a) and comparing with the baseline bedroom measurement in Figure16a, no additional peaks appear in the EQS spectrum observed using OSC, WRG or WR. All the peaks that we do observe in Figure17a (@64kHz, 101kHz, 122kHz, 202kHz and 214kHz) were also present in baseline measurement with all appliances off (possibly coming from capacitive switching harmonics in supply) in Figure 16a, hence it can be concluded that there was no interference from the air conditioner. On performing similar analysis on other listed appliances, it was observed that the appliances that create peaks in the EQS region include CFL lights (Figure 17c, e & h @40-60kHz), fluorescent tube lights (Figure 17f, m & n @40-60kHz) and laptop adaptors (Figure 17o @90-110kHz ), while all other appliances including LED lights (Figure 17d & i), air conditioner (Figure 17a), washing machine (Figure 17j), stove (Figure 17k), fan (Figure 17b) and refrigerator (Figure 17l) do not create any interference. To get a wider perspective of our in-the-wild measurements, an open field measurement was conducted (away from power lines and with no electronic appliances nearby) using WR as shown in Figure 18. Peaks were observed @50kHz and @150kHz from the wearable device, and @95kHz and @220kHz from the laptop used for this measurement. The reason for appearance of peaks from devices is due to the power convertors (both DC/DC and AC/DC converters) employed, which are present in CFL and fluorescent lights, laptops, laptop adaptors and equipment with digital displays (in our case microwave). The switching frequencies employed by these converters appear as peaks in the spectrum. For instance, the switching frequency used by CFL and fluorescent drivers generally ranges from 40k-60kHz, to achieve optimized performance as well as to avert human visible flicker. Laptop adaptors and phone chargers both house AC/DC converters, however, since laptop chargers work at higher voltage levels (12-25V) compared to phone chargers (~5V) they couple to the body more strongly and show up as interference peaks at around 90-110kHz [20]. The peak shown from laptop adaptor is -60dBV in Figure 17o, however, this measurement is the worst-case scenario as it is conducted on an old charger. Another important note is about interference from laptops itself, due to employment of laptop's Bluetooth in recording the measurement data, it is hard to circumvent its effect in the readings and is directly affected by the laptop's closeness to the subject's body wearing the wearable receiver. Laptops themselves have in-built DC/DC converters which result in a significant interference peak even without the adaptor @90-95kHz. Though, it peaks @90-95kHz, it is broadband in nature and may raise the NF over the entire EQS spectrum in worst cases. A possible solution to avert this broadband interference is through suitable channel encoding schemes to improve the SNR performance and reduce BER. Digital displays commonly present in microwaves, stoves and refrigerators also employ AC/DC converters to power the digital circuitry. To view the effect of digital displays, we use a microwave and its effect can be seen in Figure 17g. There is a slight increase in NF in the WRG measurement when microwave is turned on, however, no visible effect is observed in the WR measurement as the coupling from it is negligible due to smaller surface area (as compared to light from lamps and tube lights). Another interesting observation was made in the lab location having dimmable fluorescent tube lights (Figure 17m & n). The lights showed two different frequencies at two different light settings, namely 40kHz and 53kHz attributed to its frequency-based feedback mechanism for changing the light intensity [21]. It is to be noted that while CFLs and fluorescent lights cause interference peaks, LED lights do not. This is because LED drivers employ a much lower switching frequency of the order of 100's of Hz, hence not visible in the spectrum [22]. Also, additional peaks due to buffer nonlinearities show up while taking CFL and fluorescent light measurements (Figure 17c, e, f & h) due to the high magnitude of interference peak recorded (~50dBV @44kHz) giving rise to $2^{nd}$ order peak @88kHz, $3^{rd}$ order peak @132kHz and so on. All other appliances (stoves, refrigerators, fans, air conditioners) work on AC supply and show a peak at 60Hz (seen as a raised peak at DC in the measurements which is also due to the coupling of 60Hz main AC supply from house wiring to the body). In summary, we can say that the largest impediment in the EQS region are interferences from lights and laptops which must be accounted for while using EQS HBC.

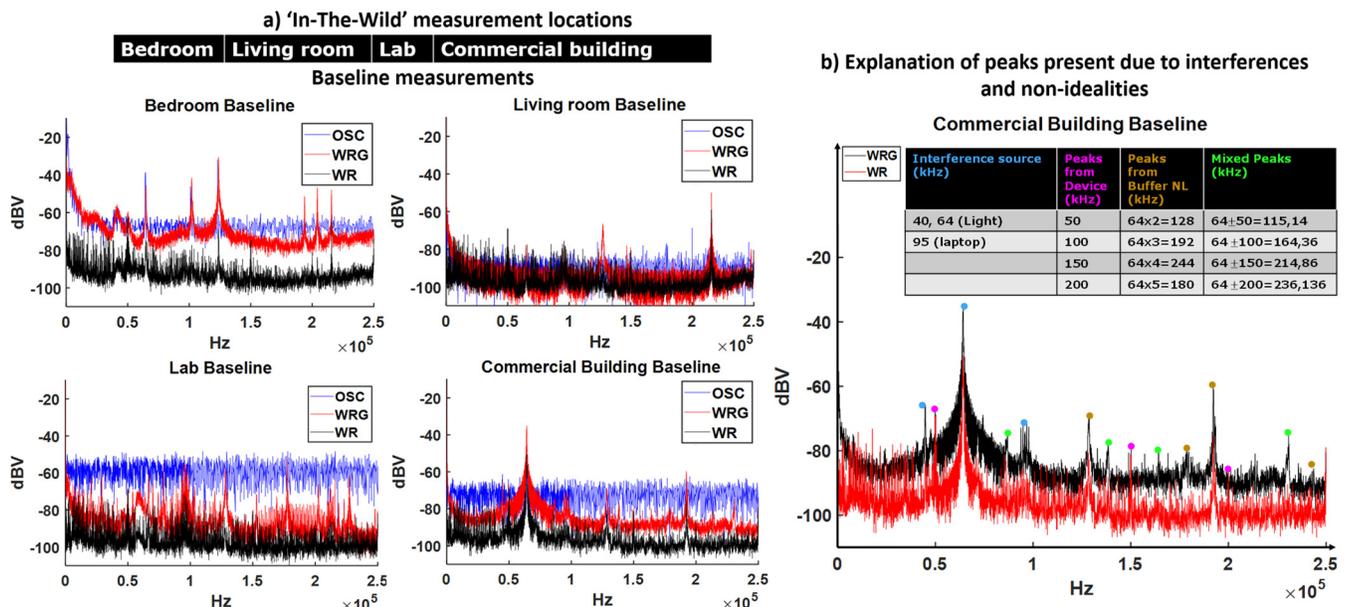

Figure 16. a) Measurement of baselines at different locations using OSC, WRG and WR. b) Explanation of the peaks seen in the commercial building hall baseline.

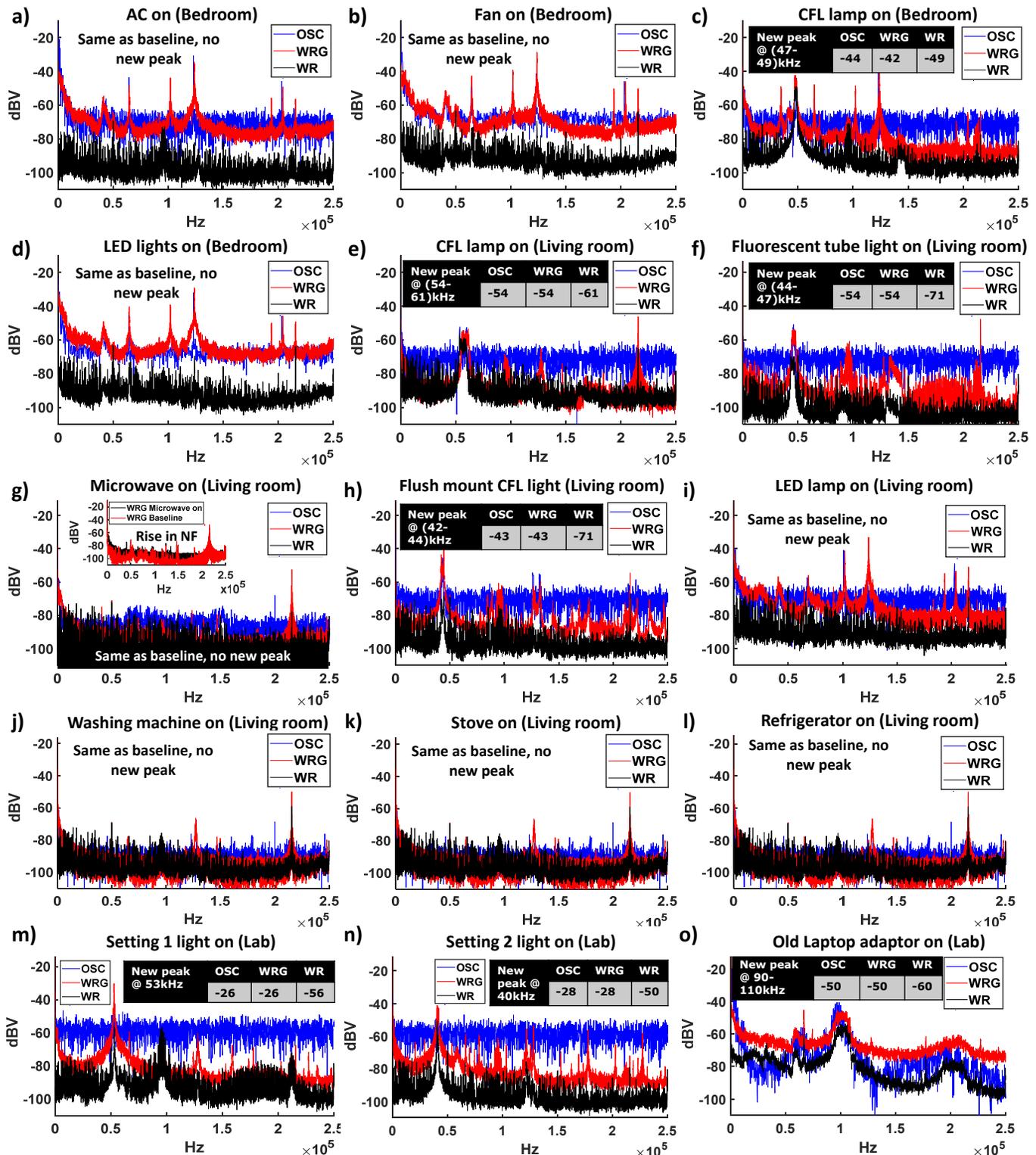

Figure 17. Measurement of the effect of individual appliances in the EQS spectrum using OSC, WRG and WR at different locations.

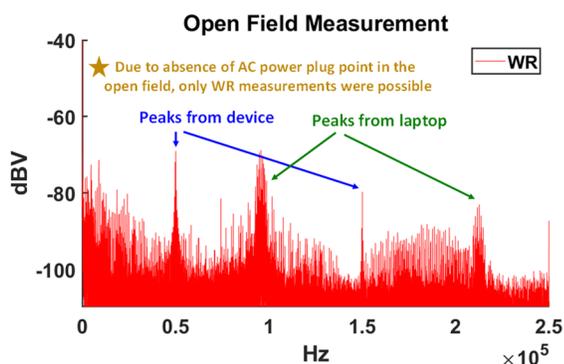

Figure 18. Interference measurement in the EQS spectrum conducted in an open field using a wearable measurement device.

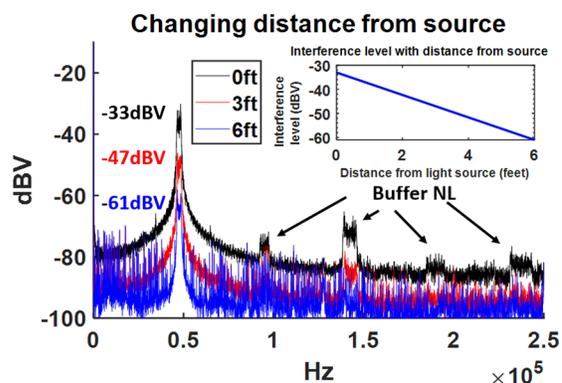

Figure 19. Effect on received interfernce signal on WR on changing distance from source, or in other words varying $C_{INTF}$.

With this understanding of the possible interference sources, we can now try to explain the peaks in the commercial building hall baseline (Figure 16b). A significant peak is present around 64kHz in WRG, which from our understanding can be attributed to the fluorescent lights present in the hall. Another observation is that the signal strength is quite high (~ -38dBV) hence it is likely that second order products of mixing and buffer nonlinearities will be present. On checking for peaks from mixing effect between the interference from fluorescent lights (64kHz) and from the measurement device (50kHz, 100kHz, 150kHz &200kHz) as discussed in section VIC, peaks at 164kHz (=64kHz + 100kHz) and 236kHz (= 64kHz + 200kHz = 264kHz aliased to 236kHz) were observed. Similarly, on checking for peaks from buffer nonlinearity at integral multiples of 64kHz, peaks at 128kHz (from $2^{nd}$ order buffer nonlinearity = 2x64 kHz), at 192kHz ($3^{rd}$ order), at 244kHz ($4^{th}$ order) and at 180kHz ($5^{th}$ order) were observed. Additionally, an interference peak from laptop was observed @95kHz. In summary, all the peaks in the commercial building hall can be explained to be coming from fluorescent lights and the laptop adaptor. This also reaffirms our statement that typically the major source of interference in any setting is from lights and must be taken care of during EQS HBC.

The effect of distance from interference source on the magnitude of received signal by the WR is shown in Figure 19. This change in received signal with distance from source is due to change in $C_{INTF}$ value. At about 2 inches from the CFL light bulb, the interference magnitude picked up by the WR is -33dBV, -47dBV at ~3feet from source and -61dBV at ~6feet from source. Thus, it can be noted that the magnitude of interference picked up is strongly dependent on the closeness to the source. In other words, the reduction in distance between the source and body, increases $C_{INTF}$ value which in turn enhances our received signal as explained in Figure 5.

It is possible to avert the undesirable effects from device, mixing and buffer nonlinearities through improved receiver design. Under such circumstance, the EQS HBC band would be as shown in Figure 20 which combines the observed peaks from our measurement results to better visualize the EQS spectrum and to choose the appropriate frequencies for communication, including the peaks from lights (40k-70kHz) and from the laptop (90-100kHz). It must be noted that the peaks present in the baselines (Figure 16a) have not been included in Figure 20 (e.g. @215kHz), as they are location dependent. To this end, it may be useful for future EQS HBC devices to incorporate techniques that will take a decision on the choice of band real-time based on the detected interferences. However, coming back to the current in-the-wild measurements, Figure 20 gives us an inkling about how the interferences as received by an actual EQS HBC receiver would be like. This is extremely useful in deciding for the proper choice of band for communication avoiding any potential interference regions. The typical EQS HBC received signal value ranges from ~-60 to -75dBV. The worst-case SIR of ~-10 to -25dB occurs below 100kHz due to interference from light and is seen to improve beyond it, along with a sufficiently high SNR of ~15 to 30dB. Hence, the ideal choice for EQS HBC would be frequencies beyond 100kHz where the possibility of any interference is less.

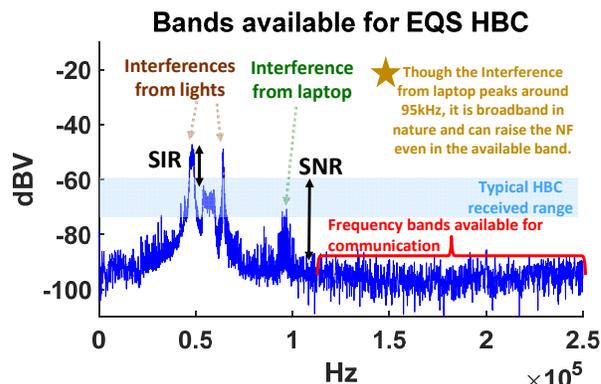

Figure 20. Viewing all interferences from sources combined to determine the available bands for EQS HBC.

## VIII. Conclusion

A proper choice of frequency band is needed for interference free EQS HBC as it will affect any data transmission through the body. The measurements conducted using oscilloscopes or spectrum/network analyzers as receiving equipment, does not accurately represent the actual wearable scenario in the EQS HBC frequency range (few MHz). This necessitates the use of a battery-operated measurement device with a small ground plane and high capacitive input termination. Thus, a miniaturized wearable sized measurement device is designed which accurately recreates the actual wearable scenario and for the first time in-the-wild interference measurements using a wearable device are presented showing interferences from large surface area radiating devices such as power supply lines, switching circuitry in lights sources, digital displays, laptops and laptop adaptors which get capacitively coupled to the body and pose issues during communication. Furthermore, a human biophysical interference pickup model is proposed, and interference measurement results are analyzed allowing for a proper choice of band for interference-free operation beyond 100kHz.

## Acknowledgement

This work was supported in part by Eli Lilly through the Connected Health care initiative, Air Force Office of Scientific Research YIP Award under Grant FA9550-17-1-0450 and the National Science Foundation Career Award under Grant CCSS 1944602.